\documentclass[12pt]{article}
\usepackage{amsmath}
\usepackage{graphicx}
\usepackage{enumerate}
\usepackage[numbers]{natbib}
\usepackage{amsmath}
\usepackage{amssymb}
\usepackage{amsthm}
\usepackage{bbm}
\usepackage{algorithm}
\usepackage{algpseudocode}
\newtheorem{proposition}{Proposition}
\newtheorem{assumption}{Assumption}
\newtheorem{remark}{Remark}
\usepackage{url} %
\usepackage[T1]{fontenc}
\usepackage[T1]{fontenc}

\usepackage[colorlinks=true,
            linkcolor=blue,
            citecolor=blue,
            urlcolor=blue]{hyperref}

\newcommand{\mH}{{\mathcal{H}}}

\newcommand{\mL}{{\mathcal{L}}}

\newcommand{\mP}{{\mathcal{P}}}

\newcommand{\mS}{{\mathcal{S}}}

\begin{document}

\def\spacingset#1{\renewcommand{\baselinestretch}%
{#1}\small\normalsize} \spacingset{1}

{
  \title{\bf %
  Precision Physical Activity Prescription via Reinforcement Learning for Functional Actions}
  \author{Gefei Lin\hspace{.2cm}\\
    Department of Statistics, The George Washington University\\
    Rui Miao \\
    Department of Mathematical Sciences, The University of Texas at Dallas \\
    Jennifer Sacheck \\
    Department of Behavioral and Social Sciences, Brown University\\
    and\\
    Xiaoke Zhang\\
    Department of Statistics, The George Washington University}
  \maketitle
} 

\newcommand{\publiccodeurl}{https://github.com/gefeilin/Functional-Fitted-Q-Learning}

\newcommand{\codestatement}{%
\url{\publiccodeurl}
}

\bigskip
\begin{abstract}
Physical activity (PA) plays an important role in maintaining and improving health. Daily steps have been a key PA measure that is easily accessible with common wearable devices. However, 
methods are lacking to recommend a personalized optimal distribution of daily steps over a period of time for the best of certain health biomarkers. In this paper, we fill this void based on the data from the All of Us Research Program which includes months of step counts as well as repeated measurements of key health biomarkers. 
We 
develop a new offline reinforcement learning (RL) algorithm to learn personalized and optimal PA distributions associated with cardiometabolic risk, where the action is a function representing the daily step distribution over a period of time. Simulation studies demonstrate the advantage of 
the proposed approach 
over existing continuous-action RL methods. 
The learned optimal policy from the All of Us data generally suggests people take more daily steps and also follow a more consistent pattern of PA over time while offering tailored recommendations for subgroups in blood glucose level, body mass index, blood pressure, age, and sex.

\end{abstract}

\noindent%
\vfill

\newpage
\spacingset{1.9} %
\section{Introduction} \label{sec:intro}

Physical activity (PA) is known to be important for multiple aspects of human health, such as cardiovascular and cardiometabolic health \cite{Cleven2020PAHealth}, cognitive function \cite{Barreto2016PACognition},
muscular strength \cite{Rostron2021PAMuscle}, and the risk of early mortality \cite{delPozoCruz2021LightPA}.
Despite its well-documented health benefits, the World Health Organization (WHO) reports that a large proportion of the population remains insufficiently active, with about 31\% of adults worldwide failing to meet recommended levels \citep{Strain2024LancetGlobalHealth,WHO2024physicalactivity}.
 While both the U.S. Physical Activity Guidelines for Americans \cite{USDHHS2018PAguidelines} and the WHO \cite{WHO2020PAguidelines} provide recommendations primarily in terms of PA intensity (e.g., minutes of moderate-to-vigorous activity assessed by accelerometry or self-report questionnaires), daily step count has emerged as a particularly attractive metric. Daily steps can be easily captured by wearable devices, such as pedometers and accelerometers, which are practical for collecting data over long periods and intuitive for both individuals and clinicians to interpret \cite{Kraus2019DailySteps}.

A growing body of evidence demonstrates that daily steps are strongly associated with diverse health benefits \cite[e.g.,][]{Ahmadi2024StepsMortalityCVD,delPozoCruz2022StepsDementia,Paluch2022DailySteps}. However, traditional analyses in health studies typically reduce steps to highly summarized PA measures, which can obscure critical behavioral patterns. For instance, Tudor-Locke et al. \cite{TudorLocke2004StepsEnough,Tudor-Locke2008RevisitingSteps} translated existing PA guidelines into mean daily step equivalents and established a categorization of PA (i.e., sedentary, low active, somewhat active, active, and highly active) based on mean daily steps. While influential, such scalar or categorical summaries collapse the full distribution of PA and overlook heterogeneity across individuals---for example, the same level of mean steps may have different health impacts depending on how activity is %
distributed 
or on individual characteristics such as age, sex, or metabolic status.

To address these limitations, recent studies have begun to explore the distributional representations of PA. The PA distribution can not only avoid critical information loss as by scalar summaries but also address the misalignment issue across subjects (e.g., as in Figure \ref{fig: example of aou}).
Ghosal et al. \cite{Ghosal2025OutcomeRegression} 
developed 
regression frameworks based on quantile functions
and showed that modeling distributional PA improves predictive accuracy for digital health outcomes compared to scalar summaries. 
Matabuena and Petersen \cite{MatabuenaPetersen2023} proposed a distributional representation of accelerometry data in the US National Health and Nutrition Examination Survey (NHANES) 
and extended nonparametric survey regression models, showing improved associations with biomarkers and mortality risk compared to traditional scalar metrics. 
Wang et al. \cite{Wang2023FlexibleFunctionalTE}
developed a flexible 
functional treatment effect estimation framework and applied it to the NHANES accelerometry data to study the causal effect of the PA distribution. 
Long and Zhang \cite{Long2025CausalPA} re-examined the causal effect of the PA distribution based on the NHANES accelerometry data in the presence of unmeasured confounding. 
All these studies above have demonstrated the benefits of treating PA exposures as distributional objects rather than scalar summaries. 
However, 
they have only focused on estimating the associations or causal effects of distributional PA on health outcomes. To the best of our knowledge, no method has been available to provide ``precision PA prescriptions'', that is individualized PA recommendations based on a person's covariates such as %
age, sex, and key health biomarkers.
In this paper, we will %
address this gap by proposing %
a distribution-valued precision PA prescription framework, specifically in the form of a conditional distribution given key covariates.

The All of Us Research Program \cite{AllofUsInvestigators2019}, formerly the Precision Medicine Initiative Cohort Program, is a large-scale NIH initiative designed to build a diverse and longitudinal health database. Participants contribute electronic health records, physical measurements, surveys, biospecimens, and wearable device data, including daily step counts from Fitbit as a measure of PA. The program's scale, diversity, and repeated measurements over time provide unique opportunities to capture dynamic changes in behavior and health, enabling the study of heterogeneous treatment effects and personalized interventions. Compared with the short monitoring window and cross-sectional design of the NHANES accelerometry data, the All of Us program is better positioned to support the development of adaptive and data-driven strategies. Therefore, the practical objective of this paper is to find the optimal precision PA prescriptions for specific cardiometabolic biomarkers with consideration of sex and age based on the All of Us daily step count data.

To address this problem quantitatively, we adopt the reinforcement learning (RL) framework. 
RL is a branch of machine learning that studies how an agent interacts with an environment to make sequential decisions. At each time step, the agent observes the current state, selects an action according to a policy (e.g., strategy or PA prescription), and receives feedback in the form of a reward. Through repeated interaction, RL algorithms learn to optimize a policy that maximizes the cumulative reward, thereby balancing immediate benefits with long-term outcomes. RL has been applied in PA, primarily to design just-in-time adaptive interventions%
, where algorithms decide whether 
to deliver context-tailored reminders for 
activity 
through mobile apps \cite{Liao2020HeartSteps}. Thus their focus is substantially different from ours, which is to recommend the optimal PA distribution for an individual to follow.

Prior work has provided a proof of concept that functional policies are feasible within RL where actions are modeled as functions over a domain. For example, Pan et al. \cite{pan2018functionvalued} studied partial differential equation control in an online RL setting, using descriptors and adapters to parameterize functional actions and refining policies through repeated interaction with the environment. However, 
such methods  %
have not yet been applied to pre-collected observational health data%
, where interpretability
and data efficiency are crucial and 
trial-and-error interactions are usually infeasible due to ethical constraints.

In this paper, 
we develop a new offline RL algorithm for functional policy learning 
relying solely on pre-collected data where the action is a function. 
 We extend the classic fitted-Q iteration algorithm \cite{ernst2005tree} 
to incorporate the functional action. To ensure the smoothness of the optimal functional policy, we propose to iteratively update it by penalized splines. 
A simulation shows that our proposed method for the functional action is consistently superior over state-of-the-art offline continuous-action RL algorithms  which aggregate the functional action into a continuous action. We apply the proposed method 
to the All of Us step count data 
to learn the optimal and interpretable distribution-valued PA prescription for the best of cardiometabolic health.

The main contributions of this paper are threefold.
First, to the best of our knowledge, this paper is the first attempt to learn precision PA prescriptions. 
Instead of summarizing PA measurements substantially, we represent PA prescriptions as distributions of daily steps to preserve the full spectrum of different daily step count levels. The precision PA prescription we aim to recommend is the distribution of daily steps given an individual's covariates. 
Second, a new offline RL algorithm is developed that incorporates a smooth functional action and outperforms state-of-the-art algorithms for continuous actions. Lastly, the analysis of the All of Us step count data provides the first 
precision PA prescription %
with respect to cardiometabolic risk and also reveals how covariates influence the recommended PA distributions.

The remainder of the paper is organized as follows. 
Section~\ref{sec:allofus} describes the All of Us data. 
Section~\ref{sec:method} presents the proposed methodology. 
Section~\ref{sec:pa_prescription} provides the PA prescription results based on the All of Us data. 
Section~\ref{sec:discussion} concludes the paper with a discussion and future work. 
The code for this paper is available at \codestatement. An interactive CardioStepRx web app is available at \url{https://ling2-cardiostep-rx2.share.connect.posit.cloud/}.

\section{All of Us PA Data} \label{sec:allofus}

\begin{figure}[!h]
    \centering
    \includegraphics[width=1\linewidth]{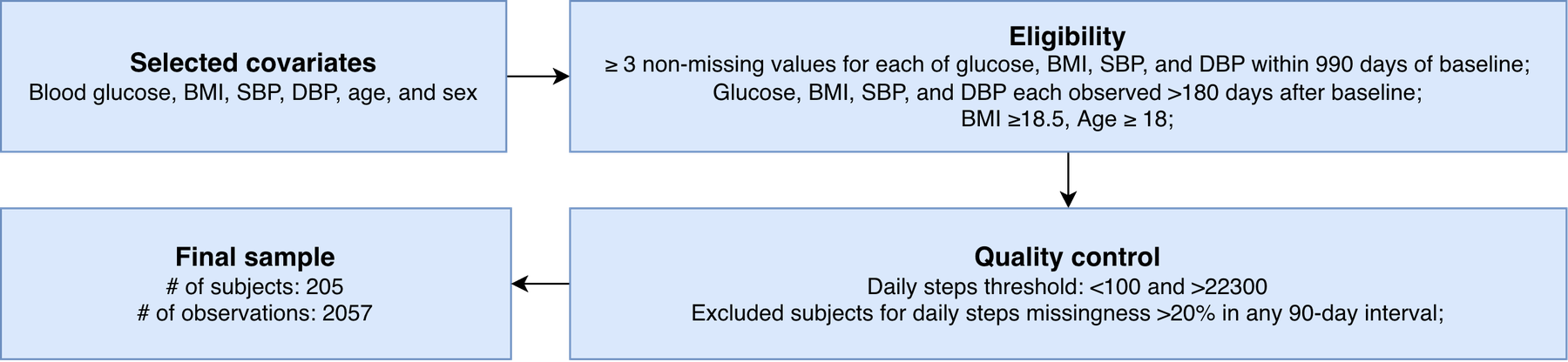}
    \caption{Data construction outline.}
    \label{fig:data_outline}

\end{figure}

We %
analyzed a longitudinal PA dataset constructed from the All of Us Research Program \cite{AllofUsInvestigators2019}, which %
collected electronic health records, surveys, physical measurements, biospecimens, and wearable device data.
Daily step counts derived from Fitbit device records were used as the PA measure. 
Fitbit devices count steps using a three-axis accelerometer. 
Fitbit data in the All of Us Research Program are generated by the Fitbit Web API and contributed by participants who consent to share or sync their Fitbit data through the All of Us Participant Portal or through the Wearables Enhancing All of Us Research (WEAR) Study.
 To link PA with %
 cardiometabolic health biomarkers, we incorporated repeated measurements of blood glucose, body mass index (BMI), systolic blood pressure (SBP), and diastolic blood pressure (DBP), along with demographic covariates including sex and age.
 The data construction is illustrated in Figure~\ref{fig:data_outline}.

Explicitly, %
the baseline time was operationally defined as the date of the first available glucose measurement, which is the only laboratory-based measurement among the cardiometabolic biomarkers in this analysis.
For each participant, we extracted a $990$-day observation window starting from baseline and divided it into consecutive $90$-day intervals. To ensure sufficient longitudinal information, we required participants to have at least three measurements of blood glucose, BMI, SBP, and DBP during the observation window, and also to have daily step counts available across the same period. Because biomarker measurements were not always available exactly at the $90$-day boundaries, values were aligned to each interval using the last observation carried forward (LOCF) \cite{fleming2019missingness, aguiar2020exploring}, 
providing consistent coverage across participants. When multiple measurements of the same biomarker occurred on the same day, we used their daily average.

We %
removed outliers and unreasonable step count values in two stages. Following the All of Us official document and prior literature \cite{AllofUs2025FitbitResources,Jeong2025AoUActivityInequality}, we initially treated days with fewer than $100$ steps/day or greater than $50{,}000$ steps/day as implausible and set them to missing. To improve the reliability of downstream distribution-valued PA recommendations, we then applied an interquartile range (IQR)-based outlier detection procedure to the step counts and treated values exceeding $22{,}300$ steps/day as outliers; these observations were also set to missing. 
Eventually the final sample consisted of $205$ participants, with 3 to 11 repeated measurements per participant. 

Among the $205$ participants, 
$149$ (72.7\%) were female while $56$ (27.3\%) were male. 
For the continuous covariates and daily steps, the mean (SD) values were 102.95 (28.80) mg/dL for glucose, 29.21 (5.74) kg/m$^2$ for BMI, 123.87 (15.15) mmHg for SBP, 74.59 (9.35) mmHg for DBP, 55.46 (14.59) years for age, and 8,314.61 (4,389.41) steps/day for daily activity. Their other summary statistics are given in Table~S1 in the Supplementary Material.

Figure~\ref{fig: example of aou} illustrates %
step count trajectories for two participants over a nine-month period. Obviously they are not temporally aligned across individuals. We thus choose to take the distribution of daily steps as the action. Our goal is to learn the optimal precision prescription of daily steps for the next 90 days, which is in the form of a conditional distribution of daily steps given covariates associated with cardiometabolic risk.
Details of the analysis will be deferred to Section \ref{sec:pa_prescription}.

\begin{figure}
    \centering
    \includegraphics[width=0.7\linewidth]{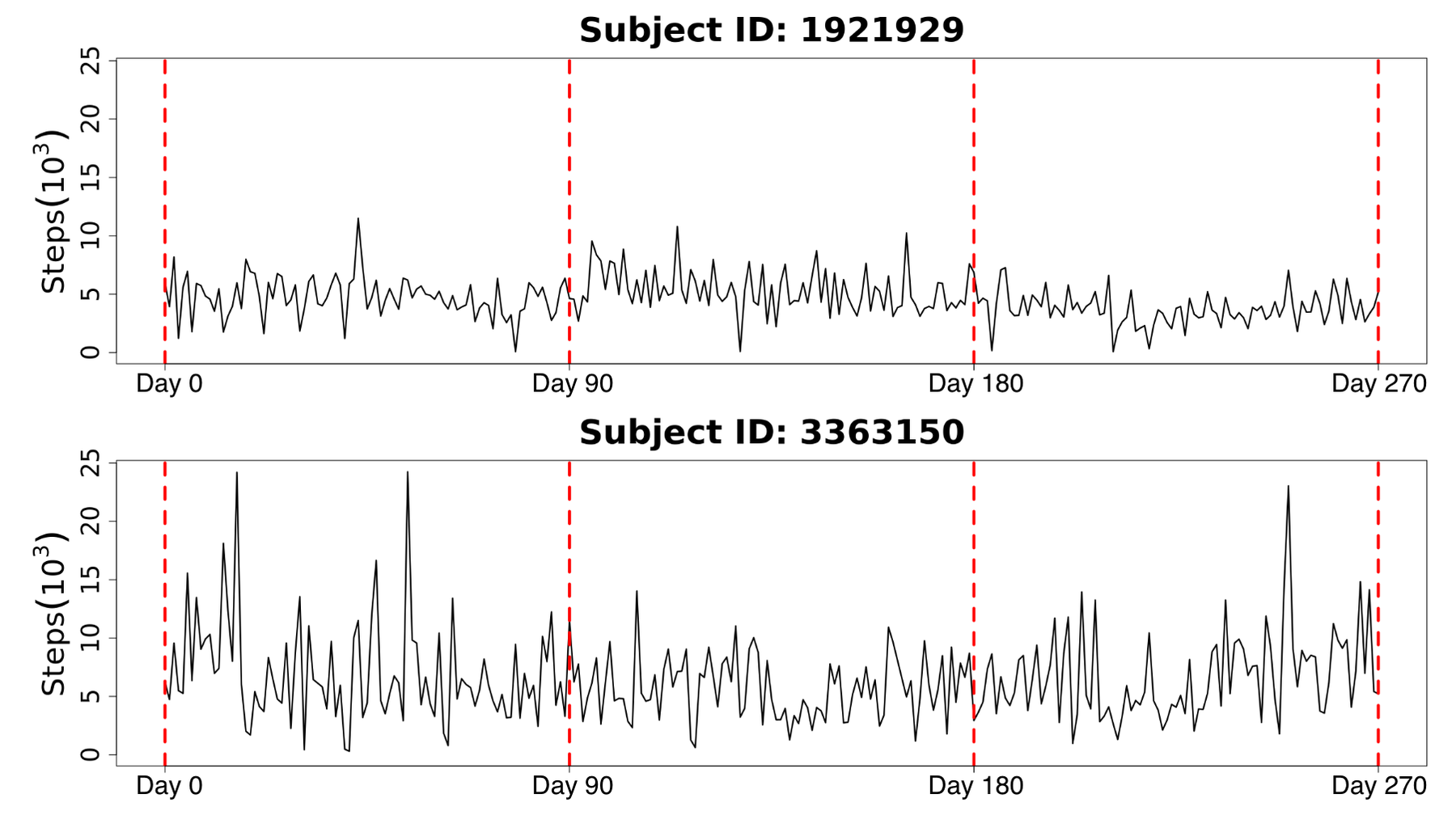}
    \caption{Daily steps of two subjects from the All of Us data. }
    \label{fig: example of aou}
\end{figure}

\section{Offline RL for Functional Actions} \label{sec:method}

In this section, we develop new offline RL algorithms where the action is a general smooth function. They will be applicable to find the optimal precision PA prescription since it is in the form of a distribution, a special type of function. 

\subsection{Problem Settings}\label{sec:prbelm_setting}

We consider a discounted infinite-horizon Markov decision process (MDP) defined by the tuple
$(\mathcal{S}, \rho, \mathcal{A}, P, r, \gamma)$,
where $\mathcal{S}$ denotes the state space,  
$\rho$ is the initial state distribution on $\mathcal{S}$, 
$\mathcal{A}$ is the functional action space,  %
$P: \mathcal{S}\times\mathcal{A}\rightarrow\mathcal{P}(\mathcal{S})$ is the transition probability operator, with $\mP(\mS)$ the space of probability measures on $\mS$, 
$r: \mathcal{S}\times\mathcal{A}\rightarrow[0,1]$ is the reward function, 
and $\gamma\in[0,1)$ is a discount factor. For simplicity and without loss of generality, we let the action space $\mathcal{A} = \mL^2([0,1])$, the functional $L^2$ space over the domain $[0,1]$.

At each discrete time step $t=0,1,\ldots$, the data are observed as a tuple $(S_t,A_t,R_t, S_{t+1})$, 
where $S_t\in\mathcal{S}$ denotes the state at time $t$, 
$A_t\in\mathcal{A}$ denotes the action taken at time $t$ according to a stationary policy $\pi$, 
and $R_t = r(S_t,A_t)$ denotes the immediate reward received at time $t$. We assume the commonly used Markov property and stationarity as follows.

\begin{assumption}[Markov property]\label{assump:markov}
The process satisfies the Markov property that
\[
P(S_{t+1} \mid S_0, A_0, \ldots, S_t, A_t) = P(S_{t+1} \mid S_t, A_t),
\]
meaning that the next state depends only on the current state-action pair.
\end{assumption}

\begin{assumption}[Stationarity]\label{assump:time_homo}
The transition dynamics are stationary, i.e.,
\[
P(S_{t+1} \mid S_t, A_t) = P(S_{t+h+1} \mid S_{t+h}, A_{t+h}).
\quad \forall h \ge 0.
\]
Thus, transition probability operator does not 
depend on time $t$. 
\end{assumption}

Here we focus on the stationary policy $\pi: \mathcal{S} \rightarrow \mathcal{P}(\mathcal{A})$, which is a time-invariant mapping. 
If $\pi$ is deterministic, it maps each given state to a specific function $a\triangleq a(\cdot)\in\mathcal{A}$; 
if $\pi$ is stochastic, 
it is a conditional distribution on $\mathcal{A}$ given a state. Since the action is a function,  a stochastic policy $\pi$ cannot be expressed as a conditional probability density function (PDF) as for the continuous action since the PDF generally does not exist for a functional variable \citep{delaigle2010defining}. 
This distinction poses challenges for directly applying standard offline RL algorithms in functional action spaces when taking the conditional expectation with respect to the action is needed. 
To address this problem, our proposed Algorithm \ref{alg:fqe} below approximates the conditional expectation with respect to the functional action 
via Monte Carlo sampling, rather than relying on an explicit conditional PDF for the action.

For a given policy $\pi$, its Q function $Q^{\pi}(s,a)$ 
is defined by 
$$Q^\pi(s, a):=\mathbb{E}^\pi\left[\sum_{t=0}^{+\infty} \gamma^t R_t \mid S_0=s,A_0=a\right],$$
where $R_t \sim r(S_t,A_t)$, $A_t \sim \pi(\cdot \mid S_t)$, $S_t \sim P(\cdot \mid S_{t-1}, A_{t-1})$, and the $\mathbb{E}^\pi$ is the conditional expectation operator over the MDP generated by $\pi$. 
Apparently $Q^{\pi}(s,a)$ is the expected discounted total reward for $\pi$ if the initial state-action pair is $(s,a)$ and $Q^{\pi}(s,a) \in [0,1/(1-\gamma)]$ since $r\in [0,1]$.
Moreover, the value function $V^\pi(s)$ for $\pi$ given the initial state $s$ is defined by 
$$V^\pi(s) := \mathbb{E}_{A\sim \pi(\cdot \mid s)}[Q^\pi(s, A)].
$$

In practice, we observe data $(S_{i,t},A_{i,t},R_{i,t})_{1 \leq i \leq N}^{1 \leq t \leq T}$ from $N$ participants at $T$ time points. They are assumed independent %
copies of the trajectory $(S_t,A_t,R_t)^{1 \leq t \leq T}$, which starts from the initial state distribution $\rho$ and selects the next action by following a behavior policy $\beta$ 
at each time point $t$. Given the observational data, our ultimate
goal is policy optimization: find an optimal policy $\hat{\pi}$ that maximizes $v^\pi = \int_\mathcal{S} V^\pi(s) \rho(s) ds$. In the offline RL setting where interacting with the environment is not possible, off-policy evaluation is often used to estimate $v^\pi$ and compare candidate policies during learning.
Next we will introduce two new offline RL algorithms for policy evaluation and policy optimization respectively.

\subsection{Functional Policy Evaluation} \label{sec:OPE}

Under the offline RL setting in Section \ref{sec:prbelm_setting}, the goal of off-policy evaluation (OPE) is to estimate $v^\pi = \int_\mathcal{S} V^\pi(s) \rho(s) ds$ based on observed data, given a target functional policy $\pi$ and an initial state distribution $\rho$. There are mainly three types of OPE methods for discrete or continuous actions in the RL literature \citep[see reviews, e.g.,][]{uehara2022review}. Since $v^\pi = \int_\mathcal{S} \mathbb{E}_{A\sim \pi(\cdot \mid s)}[Q^\pi(s, A)] \rho(s) ds$, the first type is direct methods which directly estimate the Q function and then integrate its estimator with respect to $\rho$. The second type is importance sampling methods, which rely on the ratio between the PDFs of the target policy and behavior policy to adjust for their mismatch. The third type, double robustness method, is their combination. Since a PDF generally does not exist for a functional action \citep{delaigle2010defining}, we proposed to generalize Fitted Q-Evaluation (FQE) \citep[e.g.,][]{le2019batch}, a popular direct OPE method, to handle functional actions.

Explicitly, by Assumptions \ref{assump:markov} and \ref{assump:time_homo}, we can obtain the well-known Bellman equation: 
\begin{equation}
  Q^\pi(s, a)=\mathbb{E}^\pi\left[R_t+\gamma \cdot Q^\pi\left(S_{t+1}, A_{t+1}\right) \mid S_t=s, A_t=a\right]. \label{eq:bellman} 
\end{equation}
Motivated by the fact that the true $Q^\pi$ is the unique solution to (\ref{eq:bellman}) by the Banach fixed-point theorem \citep[see, e.g.,][]{agarwal2019reinforcement}, 
FQE iteratively solves the following optimization problem: 
$$
\hat{Q}^\pi_{m} = \arg \min_{Q^\pi} \frac{1}{NT} \sum_{i=1}^N \sum_{t=1}^T \left[R_{i,t}+\gamma \cdot \mathbb{E}_{A|S_{i,t+1}}^\pi \big(\hat{Q}^\pi_{m-1}(S_{i,t+1},A)\big)-Q^\pi(S_{i,t},A_{i,t}) \right]^2,
$$
where $\hat{Q}^\pi_{m}$ is the Q-function estimate at the $m$th iteration, $m = 1,2,\ldots$.
Based on this idea, we develop Algorithm \ref{alg:fqe} below that adapts FQE to functional policies. 

\begin{algorithm}[!h]
\caption{FQE for Functional Policy Evaluation}\label{alg:fqe}
\begin{algorithmic}[1]
\Require Dataset $\mathcal{D} = \{(S_{i,t}, A_{i,t}, R_{i,t},S_{i,{t+1}})\}$, target policy $\pi$, kernels $\kappa_S$ and $\kappa_A$, initial state distribution $\rho$, discount factor $\gamma$, and number of iterations $M$.
\State Initialize $\hat{Q}^{\pi}_{0} \gets 0$.
\For{$m= 1,\dots,M$}
    \State \textbf{Target update:} \label{alg1:target-update}for each $(S_{i,t}, A_{i,t}, R_{i,t},S_{i,t+1})$ compute 
    \[
    y_{i,t,m} = R_{i,t} + \gamma \cdot \mathbb{E}_{A \sim \pi(\cdot|S_{i,t+1}) }[\hat{Q}^{\pi}_{m-1}(S_{i,t+1},A)].
    \] 
    
    \State \textbf{Q-function update:} \label{alg1:q-update} solve
    \[
    \hat{Q}^\pi_{m} = \arg\min_{g \in \mathcal{H}} \left\{ \frac{1}{|\mathcal{D}|} \sum_{i,t} \left[g(S_{i,t}, A_{i,t}) - y_{i,t,m} \right]^2 + \lambda_{m} \|g\|^2_{\mathcal{H}} \right\},
    \]
    where $|\mathcal{D}|$ is the total number of observed transitions in $\mathcal{D}$, 
    $\mathcal{H}$ is the RKHS induced by $\kappa_S \otimes \kappa_A$ and $\lambda_{m}>0$ is a hyper-parameter. 
    
\EndFor
\State \textbf{return} $\hat{v}^{\pi} = \int_{\mathcal{S}} \mathbb{E}_{A \sim \pi(\cdot|s)} [\hat{Q}^{\pi}_M(s,A)] \rho(s)ds$. \label{alg1:integrate}
\end{algorithmic}
\end{algorithm}

In Step \ref{alg1:q-update} of Algorithm \ref{alg:fqe}, $g$ is minimized over reproducing kernel Hilbert space (RKHS) $\mH$ equipped with tensor-product kernel $\kappa_{SA}=\kappa_S\otimes \kappa_A$, where $\kappa_S$ and $\kappa_A$ are kernels defined on $\mathcal{S}$ and $\mathcal{A}$ respectively, and $||g||^2_{\mathcal{H}}$ is the squared RKHS norm of $g$. By the representer theorem \citep{kimeldorf1970correspondence}, although Step \ref{alg1:q-update} is an infinite-dimensional minimization problem, its solution has a finite-dimensional representation $\hat{Q}^{\pi}_{m}(s,a) = (NT)^{-1} \sum_{i=1}^{N}  \sum_{t=1}^{T} \hat{\alpha}_{i,t}\kappa_{SA}((s,a),(S_{i,t},A_{i,t}))$ where its unknown coefficients $\hat{\alpha}_{i,t}$ can be solved by a quadratic optimization after plugging the representation back to Step \ref{alg1:q-update}.  %
Algorithm \ref{alg:fqe} essentially approximates the Q function iteratively via kernel ridge regression (Step \ref{alg1:q-update}) and eventually estimates the value $v^\pi$ by integrating the estimated Q function (Step \ref{alg1:integrate}). 

If the target functional policy $\pi$ is stochastic, since it is not in the form of a conditional PDF, $\mathbb{E}_{A \sim \pi(\cdot|S_{i,t+1}) }[\hat{Q}^{\pi}_{m-1}(S_{i,t+1},A)]$ in Step \ref{alg1:target-update}
cannot be written as or computed via an integral of $\hat{Q}^{\pi}_{m-1}(S_{i,t+1},a)$ with respect to the conditional PDF of $a$. Therefore, by the law of large numbers, we approximate $\mathbb{E}_{A \sim \pi(\cdot|S_{i,t+1}) }[\hat{Q}^{\pi}_{m-1}(S_{i,t+1},A)]$ by 
the average of $\{ \hat{Q}^{\pi}_{m-1}(S_{i,t+1},A_l)\}_l$ where $\{ A_l\}$ are sampled from $\pi(\cdot\mid S_{i,t+1})$. For $\mathbb{E}_{A \sim \pi(\cdot|s)} [\hat{Q}^{\pi}_M(s,A)]$ in Step \ref{alg1:integrate}, we adopt the same approach to approximating it.
To finally compute $\hat{v}^\pi$, the initial state distribution $\rho$ is often chosen as the distribution of $S_0$ so the outer integral in Step \ref{alg1:integrate} can be approximated via the sample mean evaluated at observed $\{ S_{i,0}\}_i$.  

\subsection{Functional Policy Optimization} \label{sec:OPO}

Under the offline RL setting in Section \ref{sec:prbelm_setting}, the goal of policy optimization is to find an optimal $\hat{\pi}$ that maximizes the value $v^\pi$. We first establish Proposition \ref{p1} on $\hat{\pi}$.

\begin{proposition}
\label{p1}
Let $\Pi$ be the set of all non-stationary and randomized policies. Define:
$$
V^{*}(s)  :=\sup _{\pi \in \Pi} V^\pi(s),\text{ and }Q^{*}(s, a)  :=\sup _{\pi \in \Pi} Q^\pi(s, a) ,
$$
which are finite since $V^\pi(s)$ and $Q^\pi(s, a)$ are bounded between 0 and $1 /(1-\gamma)$.
There exists a stationary and deterministic policy $\pi^*$ such that for all $S \in \mathcal{S}$ and $A \in \mathcal{A}$,
$$
V^{\pi^*}(s)  =V^{*}(s), \text{ and } Q^{\pi^*}(s, a)  =Q^{*}(s, a) .
$$
We refer to such $\pi^*$ as an optimal policy.   
\end{proposition}
The proof of Proposition \ref{p1} 
is omitted since it follows that of Theorem 1.7 of \citep{agarwal2019reinforcement}.
Proposition \ref{p1} implies that 
it suffices to maximize the Q-function only over the class of stationary and deterministic policies to search for the optimal policy. Noting that 
\eqref{eq:bellman} holds for any policy, the Q-function for the optimal policy $\pi^*$ satisfies:  
\begin{equation}
  Q^*(s, a)=\mathbb{E}\left[R_t+\gamma \cdot \sup_{a^\prime} Q^*\left(S_{t+1}, a^\prime\right) \mid S_t=s, A_t=a\right]. \label{eq:optimality}  
\end{equation}
Similar to FQE and based on the fact that 
$Q^*$ is the unique solution to (\ref{eq:optimality}), 
Fitted Q-Iteration (FQI) is usually used to find $Q^*$ and thus $\pi^*$ 
\citep[e.g.,][]{ernst2005tree}. A standard FQI Algorithm 
is given 
in Section S2 of the Supplementary Material. Standard FQI involves a greedy policy update step 
which 
maximizes
the estimated $Q$-function for each sample over the action space iteratively. When the action is continuous, this step is known to be 
computationally expensive and could lead to overestimation of the Q-function 
and highly non-smooth policies in $a$ 
\citep[e.g.,][]{NIPS2007_da0d1111,Neumann2008FQI}. Based on our numerical experiments which are not provided in this paper, these issues are even more severe when the action is a function.

\begin{algorithm}[!h]
\caption{Functional Policy Optimization}\label{alg:fqi}
\begin{algorithmic}[1]
\Require Dataset $\mathcal{D} = \{(S_{i,t}, A_{i,t}, R_{i,t},S_{i,{t+1}})\}$, discount factor $\gamma$, kernels $\kappa_S$ and $\kappa_A$, and 
number of iterations $M$
\State Initialize Q-function: $\hat{Q}_0(s, a) \gets 0$.
\State Initialize functional policy: $\hat{\pi}_0(s)(u)\gets 0$.
\For{$m = 1$ to $M$}

    \State \textbf{Target update:} \label{alg2:target-update} for each $(S_{i,t}, A_{i,t}, R_{i,t},S_{i,t+1})$ compute
    \[
    y_{i,t,m} = R_{i,t} + \gamma \cdot \hat{Q}_{m-1}(S_{i,t+1}, \hat{\pi}_{m-1}(S_{i,t+1})).
    \]

    \State \textbf{Q-function update:} \label{alg2:q-update} solve
    \[
    \hat{Q}_{m} = \arg\min_{g \in \mathcal{H}} \left\{ \frac{1}{|\mathcal{D}|} \sum_{i,t} (g(S_{i,t}, A_{i,t}) - y_{i,t,m})^2 + \lambda_{m} \|g\|^2_{\mathcal{H}} \right\},
    \]
    where $|\mathcal{D}|$ is 
    the total number of observed transitions in $\mathcal{D}$, $\mathcal{H}$ is the RKHS induced by $\kappa_S \otimes \kappa_A$ and $\lambda_{m}>0$ is a hyper-parameter.

    \State \label{alg2:update}\textbf{Policy update:} solve for coefficient matrix $\boldsymbol{C} = \{c_{jk}\} \in \mathbb{R}^{p \times K}$ in
    \[
    \pi_{m}(s)(u) = \sum_{j=1}^p s^{[j]} \sum_{k=1}^K c_{jk} B_k(u)
    = \boldsymbol{B}(u)^\top \boldsymbol{C}^\top s,
    \]
    by maximizing
    \[
    \frac{1}{|\mathcal{D}|} \sum_{i,t} \hat{Q}_{m}(S_{i,t+1}, \pi_{m}(S_{i,t+1})) 
    - \eta \cdot \mathbb{E}_s\left( \int \left( [\pi_m(s)]''(u) \right)^2 du \right),
    \]
    where $\boldsymbol{B}(u)=(B_1(u), \ldots, B_K(u))^\top$ are B-spline basis functions,  $[\pi_m(s)]''(u)$ is the second derivative of $\pi_m(s)(u)$ with respect to $u$, and $\eta>0$ is a hyper-parameter. 
\EndFor
\State \Return $\hat{\pi}(s)(u) = \hat{\pi}_M(s)(u)$
\end{algorithmic}
\end{algorithm}

Hence, we propose Algorithm \ref{alg:fqi} in order to obtain an optimal functional policy and also to mitigate the aforementioned issues in standard FQI. 
Algorithm \ref{alg:fqi}, which modifies FQI, includes three main iterative steps: target update (Step \ref{alg2:target-update}), Q-function update (Step \ref{alg2:q-update}), and policy update (Step \ref{alg2:update}). The target update and policy update steps are the key differences between Algorithm \ref{alg:fqi} and standard FQI. Explicitly, standard FQI often applies a greedy approach in the target update step where $\hat{Q}_{m-1}(S_{i,t+1}, \hat{\pi}_{m-1}(S_{i,t+1}))$ in Step \ref{alg2:target-update} is replaced by $\max_{a \in \mathcal{A}} \hat{Q}_{m-1}(S_{i,t+1}, a)$ for every sample $(S_{i,t}, A_{i,t}, R_{i,t},S_{i,t+1})$. This maximization causes the issues of computational intensity and Q-function overestimation. However, these issues can be avoided in Algorithm \ref{alg:fqi} by two modifications: (1) the target update (Step \ref{alg2:target-update}) does not involve any optimization at each sample; (2) it maximizes the averaged $\hat{Q}$ over the entire dataset in Step \ref{alg2:update} instead of for each sample.

Due to the representer theorem, $\hat{Q}^{\pi}_{m}$ in Step \ref{alg2:q-update} of Algorithm \ref{alg:fqi} has the same finite-dimensional representation as that in Step \ref{alg1:q-update} of Algorithm \ref{alg:fqe}.
For easy interpretation, in Algorithm \ref{alg:fqi} we only consider the class of functional linear policies:
$$
\Pi=\left\{\pi(s)(u)=\sum_{j=1}^p \beta_j(u) s_j: \text{$\beta_j$ is twice-differentiable, $j=1,\ldots, p$} \right\}.
$$
Accordingly, in Step \ref{alg2:update} of Algorithm \ref{alg:fqi} for policy update, 
$\pi_{m}(s)(u)$ is also formulated as functional linear in $s$ where each coefficient function is a linear combination of B-spline basis functions with unknown coefficients $\boldsymbol{C}$. To ensure the smoothness of the obtained policy in $u$, we propose to solve $\boldsymbol{C}$ by regularized maximization where we maximize the averaged $\hat{Q}_m$ over the entire dataset while penalizing the roughness of $\pi_m(s)(u)$ measured by $\mathbb{E}_s\left( \int \left( [\pi_m(s)]''(u) \right)^2 du \right)$. Due to the form of $\pi_{m}(s)(u)$, we can rewrite it as 
\[
\mathbb{E}_s \left( \int \left( [\pi_m(s)]''(u) \right)^2 du \right) = \operatorname{tr} \left( \boldsymbol{C} \boldsymbol{R} \boldsymbol{C}^\top \cdot \mathbb{E}_s[ss^\top] \right),
\] 
where $\boldsymbol{R} = \int \boldsymbol{B}''(u) \boldsymbol{B}''(u)^\top du \in \mathbb{R}^{K \times K}$
with $\boldsymbol{B}''(u) = (B_1''(u), \dots, B_K''(u))^\top$. In practice, $\mathbb{E}_s[ss^\top]$ is estimated by its empirical counterpart $\hat{\Sigma}_s = |\mathcal{D}|^{-1} \sum_{i,t} S_{i,t} S_{i,t}^\top$. Here in Algorithm \ref{alg:fqi} we only consider functional linear policies, which will also be applied to the simulation and real data application. However, Algorithm \ref{alg:fqi} can be easily modified to incorporate policies of more complex forms.

\begin{remark}\label{rmk:hyperparameter}
   Hyper-parameters involved in Algorithm \ref{alg:fqe} or \ref{alg:fqi}, including kernel functions $\kappa_A$ and $\kappa_S$ and smoothing parameters $\lambda_m$ and $\eta$, are important for their empirical performances. We discuss how to properly select them in 
   Section S3 of the Supplementary Material. The hyper-parameter selection methods are shown to perform satisfactorily in the simulation study and All of Us data analysis. 
\end{remark}

\begin{remark}\label{rmk:simu}
We conducted a simulation study to assess the numerical performance of Algorithms \ref{alg:fqe} and \ref{alg:fqi} with various choices of the horizon $T$, sample size $N$, and discount factor $\gamma$ inspired by the All of Us data. 
Detailed settings and results are given in Section S4 of the Supplementary Material. 
The simulation results show that (1) Algorithm \ref{alg:fqe} can accurately estimate the value of $\pi$ of a given target functional policy;
(2) The optimal value achieved by Algorithm~\ref{alg:fqi} is consistently larger than 
those by state-of-the-art offline RL algorithms for continuous actions if the functional action after aggregation is treated as a continuous action, 
which demonstrates the benefit of using functional actions over continuous actions. 
\end{remark}

\section{Precision PA Prescription}\label{sec:pa_prescription}

In this section, we analyze the data constructed as in Section~\ref{sec:allofus} using the proposed Algorithm \ref{alg:fqi}. 
We will first learn the optimal 
distribution-valued PA policy that 
minimizes cardiometabolic risk with a long-term goal of lowering all-cause mortality risk
while %
limiting implausibly extreme PA recommendations.
In comparison with the behavior policy, the learned policy will provide precision prescriptions on how people should make adjustments on the distribution of their daily steps for the best of cardiometabolic health. 
Moreover, we will further characterize differences between the learned policy and the behavior policy across subgroups in covariates, including glucose, BMI, blood pressure, age, and sex.

\subsection{RL Formulation}\label{subsec:pa_rl_formulation}

The decision times $t=0, 1, \ldots$ are indexed in 90-day increments (i.e., Day 0, Day 90, Day 180, etc). At each time $t$, the state variable
$
S_t=(G_t,\mathrm{BMI}_t,\mathrm{SBP}_t,\mathrm{DBP}_t,\mathrm{Sex},\mathrm{Age}_t)
$
collects glucose, BMI, systolic blood pressure, diastolic blood pressure, sex, and age, and the  action $A_t$ is a function that quantifies the distribution of daily steps for the next 90-day window.

The most na\"{i}ve choice of $A_t$ is the probability density function (PDF), but its constraint that its integral over the domain equals one leads to difficulties for the  
B-spline approximations of $\pi_m$ as in Algorithm \ref{alg:fqi}. Therefore, we represented $A_t$ and thus $\pi_m$ as an unconstrained function via the 
log-quantile-density (LQD) transformation \citep{Petersen2016FDA_DensityHilbert}. 
Specifically, for each 90-day window we (i) obtain the kernel density estimate $f_t$ of the daily steps, 
(ii) define behavior $A_t$ by applying the LQD transformation to $f_t$, i.e., $A_t(\cdot)=-\log\{f_t(q_t(\cdot))\}$ where $q_t(\cdot)$ is the quantile function corresponding to $f_t$,  
(iii) learn the optimal functional policy $\hat{\pi}$ in the LQD space, and (iv) if needed, transform $\hat{\pi}$ to another space of interest, e.g., the PDF or quantile space.

We construct the immediate reward which is associated with 
cardiometabolic risk with a long-term goal of lowering all-cause mortality risk 
while %
also limits implausibly extreme PA recommendations.
Explicitly, we first define 
$\mathrm{risk}(S_t)$
as an additive function of $S_t$, 
with component functions adapted from the coefficients in the Cox proportional hazard model in \cite{Chang2017PointBasedMortalityDiabetes}:
$
\mathrm{risk}(S_t)=\psi_{\mathrm{BMI}}(\mathrm{BMI}_t)+\psi_{\mathrm{glu}}(G_t)+\psi_{\mathrm{SBP}}(\mathrm{SBP}_t)+\psi_{\mathrm{DBP}}(\mathrm{DBP}_t),
$
where
\begingroup
\small
\setlength{\abovedisplayskip}{4pt}
\setlength{\belowdisplayskip}{4pt}
\renewcommand{\arraystretch}{0.92}
\[
\psi_{\mathrm{BMI}}(\mathrm{BMI})=
\begin{cases}
0.42, & 18.5 \le \mathrm{BMI} < 22.0,\\
0.20, & 22.0 \le \mathrm{BMI} < 24.0,\\
0, & 24.0 \le \mathrm{BMI} < 30.0,\\
0.03, & \mathrm{BMI} \ge 30.0,
\end{cases}
\qquad
\psi_{\mathrm{glu}}(G)=
\begin{cases}
0.29, & G < 70,\\
0, & 70 \le G < 140,\\
0.12, & 140 \le G < 160,\\
0.21, & 160 \le G < 200,\\
0.37, & G \ge 200,
\end{cases}
\]
\[
\psi_{\mathrm{SBP}}(\mathrm{SBP})=
\begin{cases}
0.15, & \mathrm{SBP} < 110,\\
0.03, & 110 \le \mathrm{SBP} < 120,\\
0, & 120 \le \mathrm{SBP} < 170,\\
0.04, & \mathrm{SBP} \ge 170,
\end{cases}
\qquad
\psi_{\mathrm{DBP}}(\mathrm{DBP})=
\begin{cases}
0, & \mathrm{DBP} < 80,\\
0.08, & 80 \le \mathrm{DBP} < 90,\\
0.16, & \mathrm{DBP} \ge 90.
\end{cases}
\]
\endgroup
Then the immediate reward is defined as
$$
R_t \;=\; r(S_t,A_t) \;=\; 1 + \tanh\!\left(-\,\mathrm{risk}(S_t)\;-\;0.002\cdot\left(\frac{\mu_{A_t}}{1000}\right)^2\right),
$$
where $\mu_{A_t}$ is the average daily steps over 90 days implied by action $A_t$. The quadratic penalty on $\mu_{A_t}$ discourages overly extreme PA recommendations and follows common practice in reward shaping \citep[e.g.,][]{Desman2025DistributionalRLGlucose}.

In Algorithm \ref{alg:fqi}, we set the discount factor $\gamma=0.8$ due to the median effective time horizon $T_{\mathrm{eff}}=5$ in the data and the approximation $T_{\mathrm{eff}}\approx 1/(1-\gamma)$ \citep{pmlr-v28-lattimore13}.
For each coefficient function in $\pi_m$, we represent it using cubic B-spline basis functions 
with two equally spaced interior knots. 
All hyper-parameters involved in Algorithm \ref{alg:fqi} are selected as described in Section S2 of the Supplementary Material. To select the smoothing parameter $\eta$,
we use the validation set approach also described there where the ratio between the validation and training sets is 1:1.

Under this RL formulation, we apply Algorithm \ref{alg:fqi} to obtain the optimal policy $\hat{\pi}(s)(u)$ which is in the LQD domain. In Section \ref{subsec:results} below, we will compare the learned policy and behavior policy to provide insights on the recommended distributional adjustments on daily steps for the best of cardiometabolic health.

\paragraph{Checking distributional mismatch.} 

\begin{figure}[!h]
    \centering
    \includegraphics[width=0.85\linewidth]{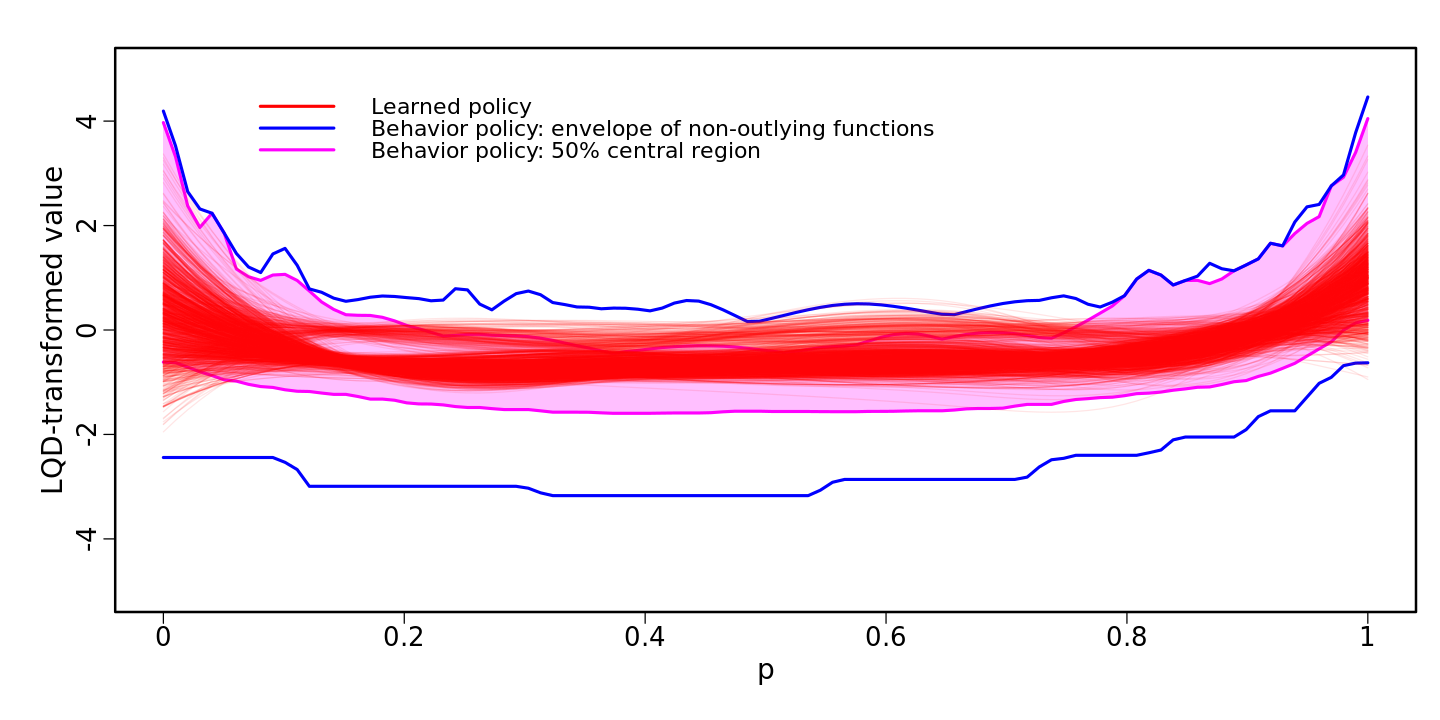}
    \vspace{-0.3in}
    \caption{Functional boxplot of learned and behavior PA distributions in the LQD domain.}
    
    \label{fig:aou_fboxplot}
\end{figure}

Before giving detailed results, we first check if there exists a substantial distributional mismatch between the optimal policy and behavior policy, which could be a critical problem in offline RL. 
After obtaining 
the learned optimal policy $\hat{\pi}(s)$ in the LQD domain, we evaluated it at all decision times, which are recommended PA distributions given their current state values, and also computed their observed PA distributions also in the LQD domain.
Then we compared them in a functional boxplot \cite{SunGenton2011FunctionalBoxplots} as illustrated in Figure~\ref{fig:aou_fboxplot}. 
Figure~\ref{fig:aou_fboxplot} shows that all the recommended PA distributions fall into the non-outlying range for the observed behavior PA distributions and most of them are even within its 50\% central region. Therefore, the learned policy is within the empirical support of the observed behaviors.

\subsection{Results and Interpretations}\label{subsec:results}

We perform two types of comparison between the learned and behavior policies, including a global comparison in the averaged daily steps and a subgroup comparison in the distribution of daily steps across covariate subgroups. 

For convenience of interpretations, we transformed the learned optimal policy $\hat{\pi}$ in the LQD domain to the quantile space, denoted by $\hat{q}(s)(p)$, which represents the recommended quantile function of daily steps over the next 90 days given the state $s$ and $p\in [0,1]$ is the quantile level. We denote the evaluated $\hat{q}$ at all decision times by $\hat{q}_{it}$. Similarly, we denote the behavior policy in the quantile domain by $\hat{q}^b$ and obtained the observed quantile function of daily steps over the next 90 days at all decision times, denoted by $\hat{q}^b_{it}$.

\subsubsection{Global comparison 
} \label{subsubsec:global}

\begin{figure}[!ht]
    \centering
    \includegraphics[width=0.55\linewidth]{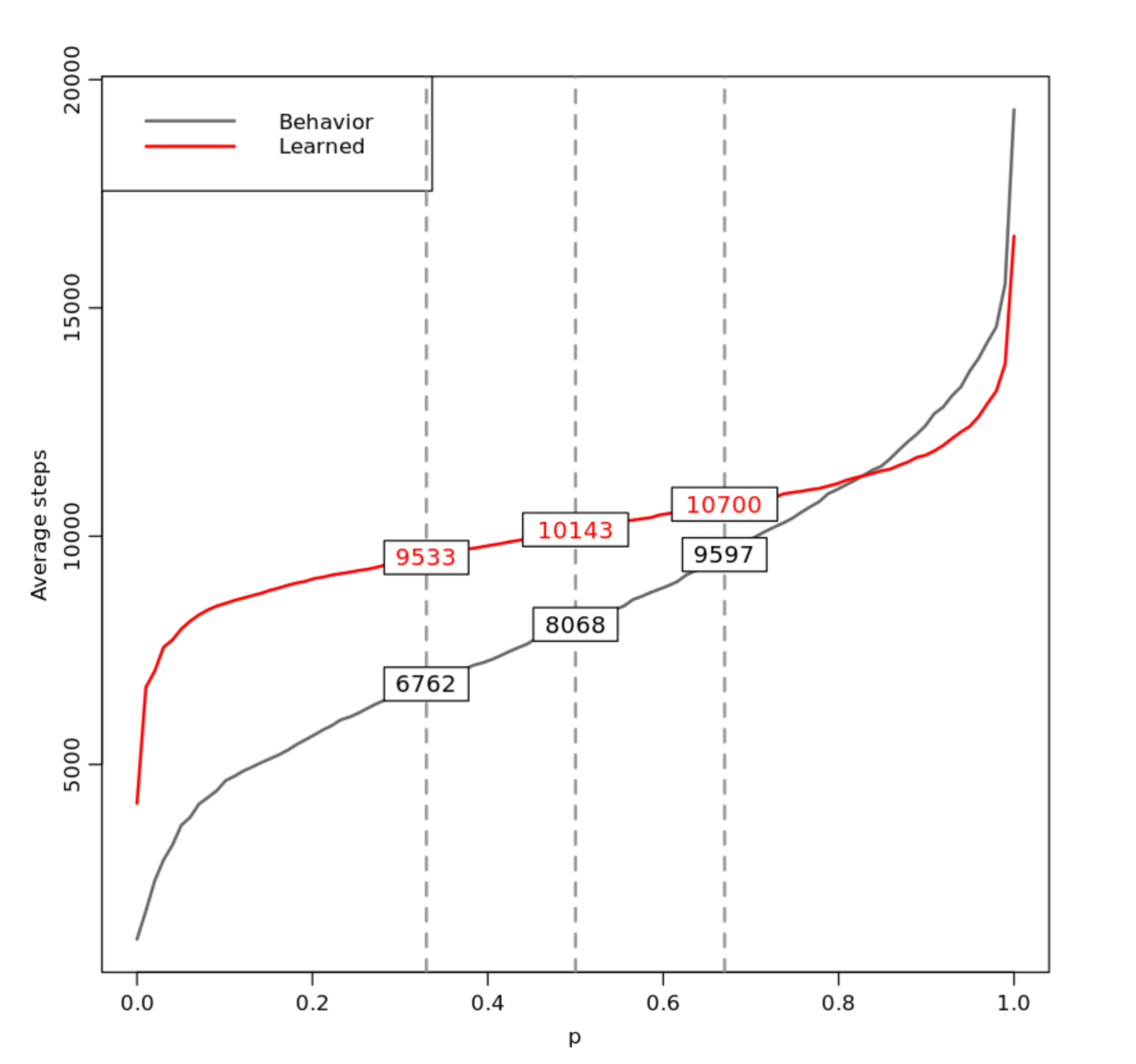}
    \vspace{-0.2in}
    \caption{Learned vs. behavioral quantile functions of 90-day average daily step counts. }
    \label{fig:aou_global_mean}
\end{figure}

Based on each $\hat{q}_{it}$ we calculated $\hat{\mu}_{it}=\int_0^1\hat{q}_{it}(p)\, dp$, which represents the recommended 90-day average daily step counts. Similarly we calculated the observed 90-day averaged daily step counts by $\hat{\mu}^b_{it}=\int_0^1\hat{q}^b_{it}(p)\, dp$. Figure~\ref{fig:aou_global_mean} illustrates the comparison between their quantile functions, where the 90-day averaged daily step counts at quantile levels $p=0.33$, $0.50$, and $0.66$ are marked. 
As shown in Figure~\ref{fig:aou_global_mean}, the quantile function for $\hat{\mu}_{it}$ is centered around approximately $10{,}000$ steps/day. Moreover, it is generally above and flatter than the quantile function for $\hat{\mu}^b_{it}$. Based on these observations, the learned policy suggests people take more daily steps in a more consistent pattern over time. 
Our proposed PA prescription is aligned with commonly used step-count targets and step-based public-health indices \citep{TudorLocke2004StepsEnough,TudorLocke2011AdultsStepsEnough}.

\subsubsection{Subgroup comparisons}\label{subsubsec:subgroup}

In this section we study the difference between the learned policy and behavior policy in each covariate. For ease of interpretations, we used the quantile forms of the learned and behavior policies, i.e., $\hat{q}(p)$ and $\hat{q}^b(p)$ respectively. 
We also discretized the continuous covariates to categorical variables following WHO-style guidelines. Explicitly, we grouped glucose into levels of Normal ($80 \le G < 120$), Borderline ($70 \le G < 80$ or $120 \le G < 150$), High ($G \ge 150$), and Low ($G < 70$), BMI into levels of Normal ($18.5 \le \text{BMI} < 25.0$), Overweight ($25.0 \le \text{BMI} < 30.0$), and Obese ($\text{BMI} \ge 30.0$), blood pressure into levels of Normal (SBP $<120$ and DBP $<80$), Elevated ($120 \le$ SBP $<130$ and DBP $<80$), and Hypertension (SBP $\ge 130$ or DBP $\ge 80$), and age into Younger (Age $<40$), Middle ($40 \leq$ Age $<60$), and Older (Age $\ge 60$). The number of observations and subjects within each level of the discretized covariate is summarized in Table S2 in the Supplementary Material.

Next we compare the difference between learned and behavior quantile functions $\hat{q}$ and $\hat{q}^b$ across different levels of each discretized covariate as well as sex.

\begin{figure}[!h]
    \centering
    \includegraphics[width=1\linewidth]{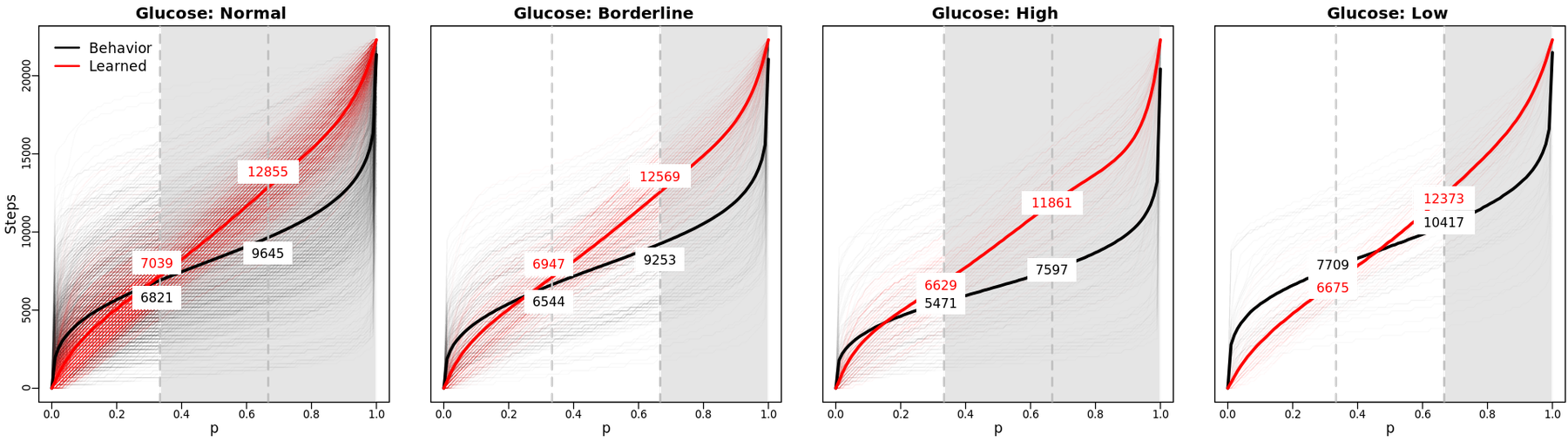}
    \caption{Learned $\hat{q}$ (red) vs behavior $\hat{q}^b$ (black) within Normal, Borderline, High, and Low glucose subgroups respectively. Solid red and black curves are averaged $\hat{q}$ and $\hat{q}^b$ respectively.
    }
    \label{fig:aou_glucose}
\end{figure}

\paragraph{Quantile difference vs Glucose.} 
The difference between $\hat{q}$ and $\hat{q}^b$ across different levels of glucose is illustrated in Figure \ref{fig:aou_glucose} where the learned $\hat{q}$ (red curves) and behavior $\hat{q}^b$ (black curves) are plotted for the observations within each of the Normal, Borderline, High, and Low glucose subgroups, together with their averages within each subgroup (solid curves). 
In each of the four plots, we included two vertical dotted lines at quantile levels $p=0.33$ and $p=0.67$ which define three activity periods: low ($0$--$0.33$), moderate ($0.33$--$0.66$), and high ($0.66$--$1$) activities, and also the daily steps corresponding to the averaged learned $\hat{q}$ and behavior $\hat{q}^b$ at $p=0.33$ and $p=0.67$. 

To select quantile intervals of $p$ at which $\hat{q}(p)$ and $\hat{q}^b(p)$ are statistically different, we applied %
the interval-wise testing 
procedure for functional data
\cite{PiniVantini2017IntervalWiseTesting} within each subgroup. 
The shaded regions in Figure \ref{fig:aou_glucose} indicate quantile intervals where $\hat{q}(p)$ and $\hat{q}^b(p)$ %
differ significantly, defined as regions where the adjusted $p$-value function was less than $\alpha = 0.05$, the interval-wise Type I error probability.

Figure \ref{fig:aou_glucose} shows that for the Normal glucose subgroup, the learned quantile function is significantly higher than the behavior one during the moderate- and high-activity periods. This indicates a practical suggestion that people of which glucose level is normal increase their steps during the moderate- and high-activity periods in the next 90 days. The suggested increase expands from about 200 steps at $p=0.33$ to about 3000 steps at $p=0.66$ and beyond until $p$ is close $1$. The patterns observed for the Borderline subgroup are similar to those for the Normal subgroup, except that the learned and behavior quantile functions are significantly different only during the high-activity period.

For the High glucose subgroup, the learned and behavior quantile functions are both lower than those in the Normal subgroup. Moreover, significant upward shifts from the observed to learned quantile function are also shown from 5,471--7,597 steps to 6,629--11,861 steps during the moderate-activity period and from $7,597$ or above to $11,861$ or above during the high-activity period. The suggested increase expands from about 1000 steps at $p=0.33$ to about 4000 steps at $p=0.66$ and beyond until $p$ is close to $1$. The recommended distributional increase in daily steps during the moderate- and high-activity periods is greater but not a lot more than that for the Normal glucose subgroup. This is probably suitable in practice: For people in the High glucose subgroup who exercise less than those in the Normal subgroup, we recommend them to increase more daily steps during the moderate- and high-activity periods for the best of their cardiometabolic health, but the suggested increase is not excessively high for them to follow.   

For the Low glucose subgroup, their behavior daily steps are already high. The learned policy only suggests a balanced adjustment on their distributions by slightly increasing their steps only during the high-activity period.

Next we apply the same procedures for the Glucose subgroups above to the subgroups in BMI, blood pressure, age, and sex respectively.

\begin{figure}[!h]
    \centering
    \includegraphics[width=0.95\linewidth]{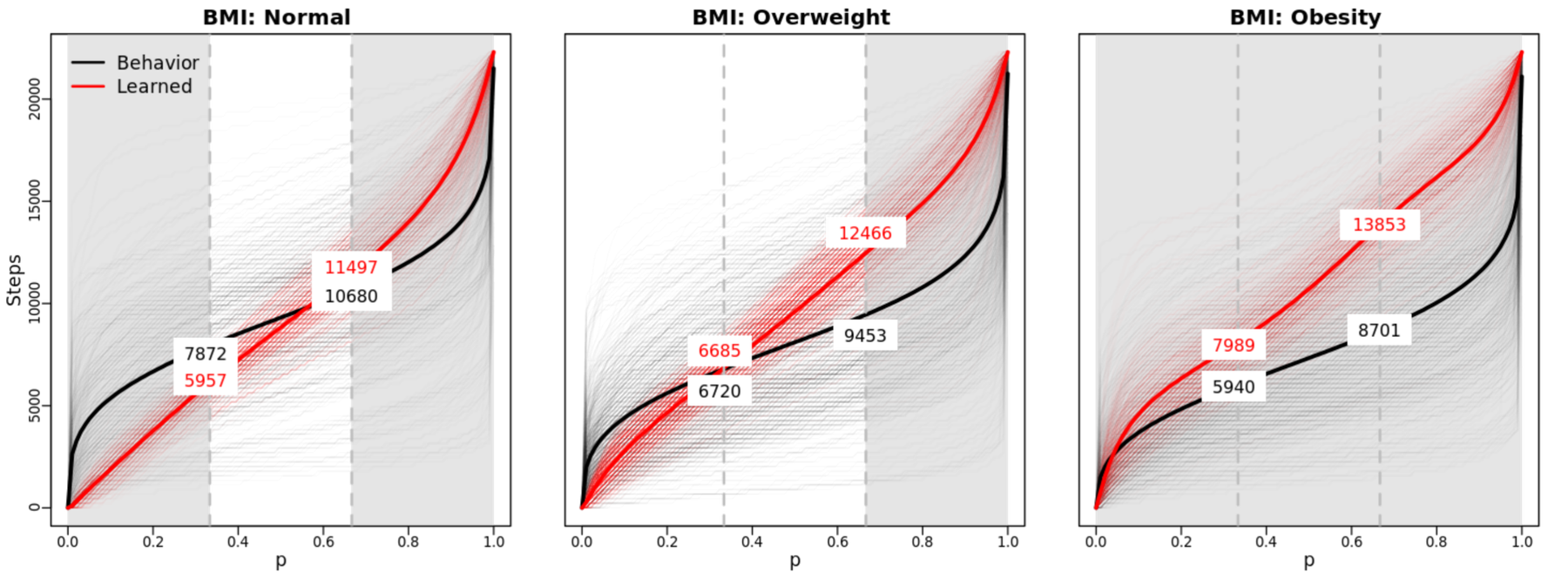}
    \vspace{-0.08in}
    \caption{Same as Figure \ref{fig:aou_glucose} except that the subgroups are the Normal, Overweight and Obese subgroups defined by BMI. 
    }
    \label{fig:aou_bmi}
\end{figure}

\paragraph{Quantile difference vs BMI.} 
As shown in Figure~\ref{fig:aou_bmi}, for the Normal BMI subgroup, the learned and behavior quantile functions are significantly different during the low- and high-activity periods. The learned policy recommended fewer steps than behavior during the low-activity period shifting from $<7,872$ to $<5,957$ steps, but more steps during the high-activity period, increasing from $>10,680$ to $>11,497$ steps.

For the Overweight group, the learned policy recommends significantly more steps than their behaviors 
only during the high-activity period, 
from $>9,453$ to $>12,466$ steps. For the Obese group, the learned policy suggests significantly more steps across all activity periods. The recommended increase ranges from about 2,000 to 5,000 steps per day, more than those recommended for the Normal and Overweight subgroups. 

\begin{figure}[H]
    \centering
    \includegraphics[width=0.9\linewidth]{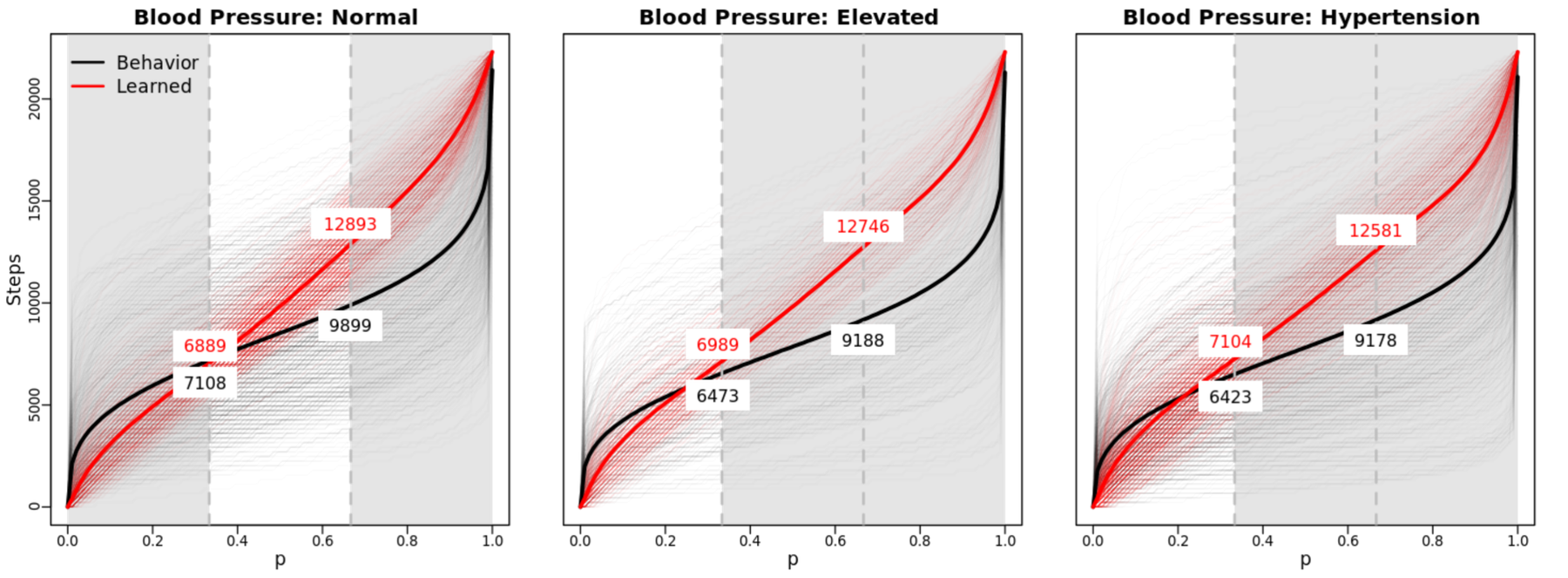}
    \vspace{-0.08in}
    \caption{Same as Figure \ref{fig:aou_glucose} except that the subgroups are the Normal, Elevated and Hypertension subgroups defined by Blood Pressure.}
    \label{fig:aou_bp}
\end{figure}

\paragraph{Quantile difference vs Blood Pressure.} 
Figure \ref{fig:aou_bp} shows that for the Normal blood pressure subgroup, the learned and behavior quantile functions are significantly different during the low- and high-activity periods. The learned policy recommends slightly fewer steps than behavior during the low-activity period 
but more steps during the high-activity period, increasing from $>9,899$ to $>12,893$ steps.

For both the Elevated and Hypertension subgroups, the learned quantile function is
significantly higher than the behavior counterpart only during the moderate- and high-activity periods, with a similar recommendation of increase about several hundreds to 3,000 steps per day.

\begin{figure}[H]
    \centering
    \includegraphics[width=1\linewidth]{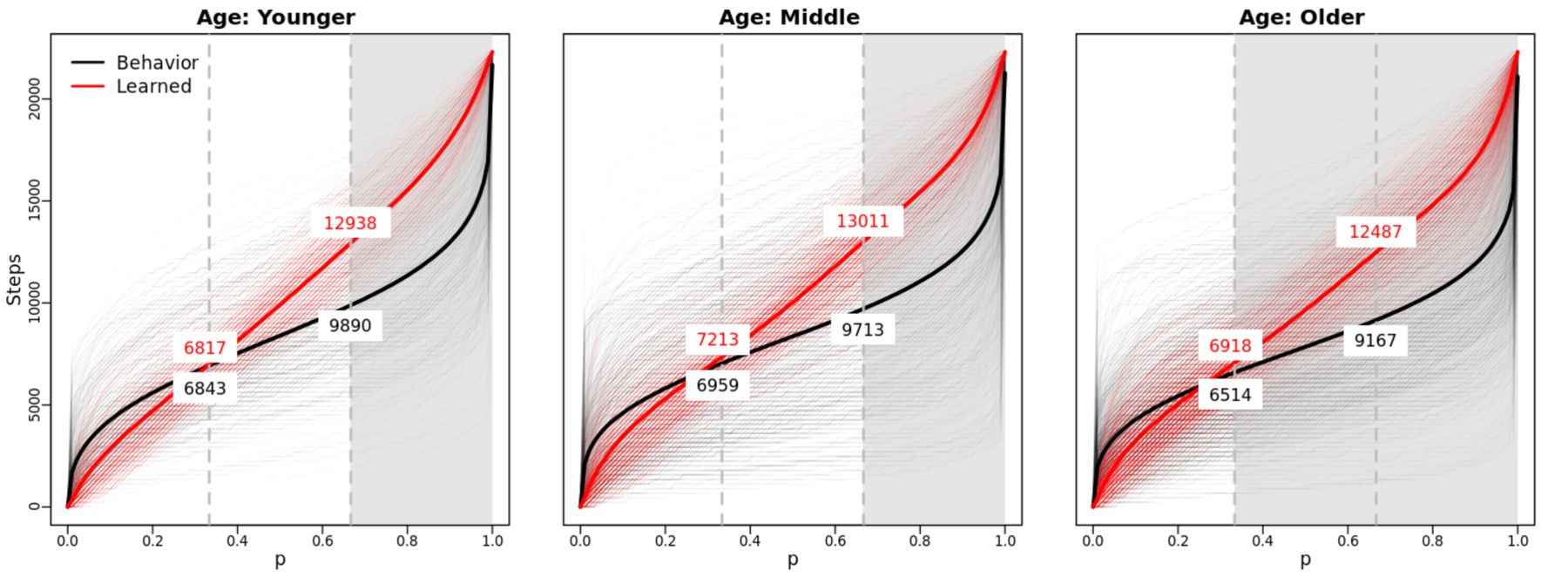}
    \caption{Same as Figure \ref{fig:aou_glucose} except that the subgroups are the Younger, Middle and Older subgroups defined by Age.}
    \label{fig:aou_age}
\end{figure}

\paragraph{Quantile difference vs Age.}
Figure \ref{fig:aou_age} shows that the learned policy recommends about 3,000 more steps during the high-activity period for all the Younger, Middle, and Older age subgroups. Key differences among the three subgroups are (1) the behavior steps for the Older subgroups are overall lower than the other two subgroups, and (2) the learned quantile function is additionally significantly higher than the behavior quantile function during the moderate-activity period.

\begin{figure}[!h]
    \centering
    \includegraphics[width=0.7\linewidth]{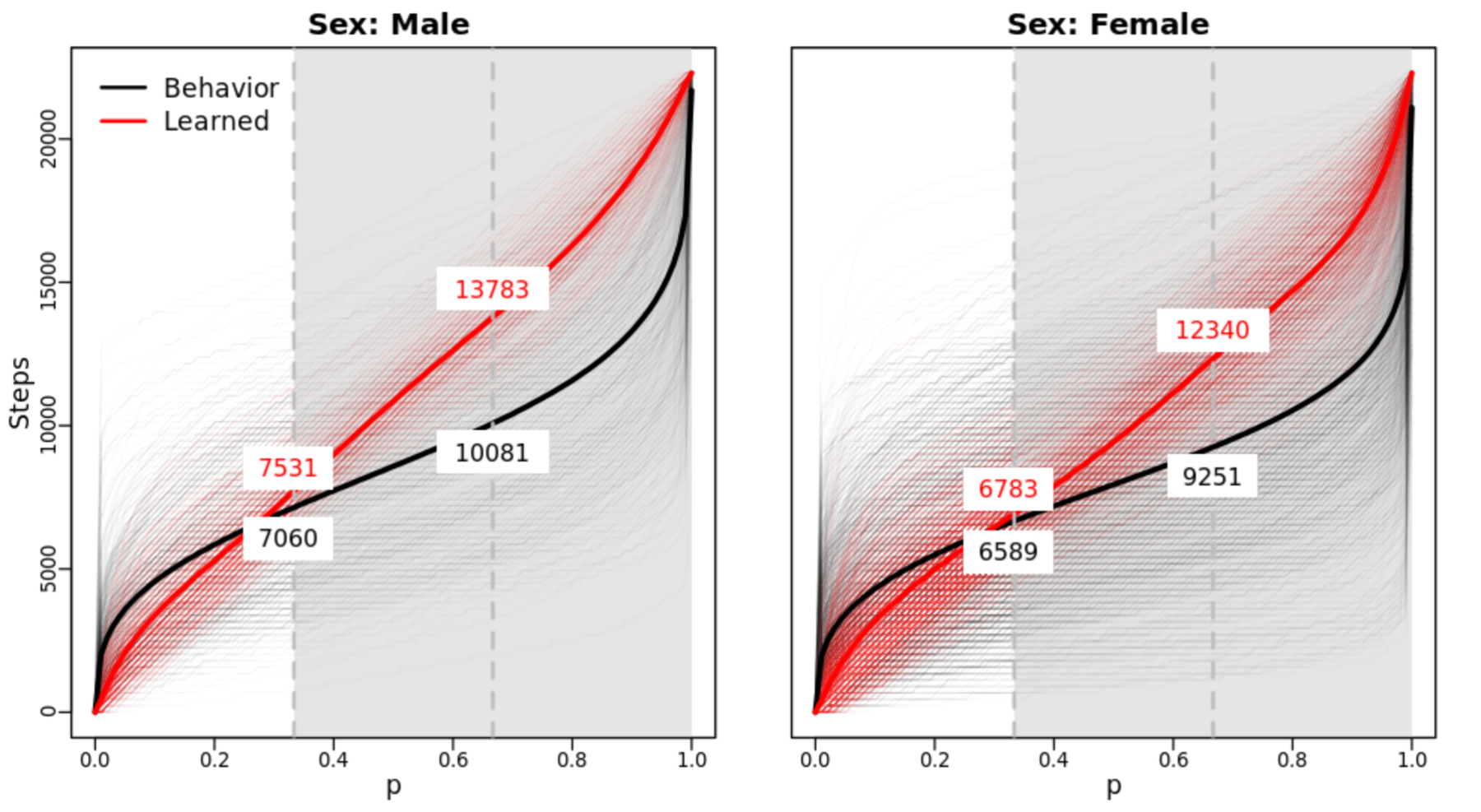}
    \caption{Same as Figure \ref{fig:aou_glucose} except that the subgroups are the Male and Female subgroups defined by Sex.}
    \label{fig:aou_sex}
\end{figure}

\paragraph{Quantile difference vs Sex.} Figure~\ref{fig:aou_sex} shows that for both males and females, the learned policy recommends significantly more steps than their behaviors during the moderate- and high-activity periods from several hundreds to 3,000 more steps, although males generally take more steps than females as revealed by their behavior quantile functions.

\subsubsection{Practical insights on precision PA prescriptions} \label{subsubsec:insights}
Here we summarize the practical insights from the results in Sections \ref{subsubsec:global} and \ref{subsubsec:subgroup} regarding precision prescriptions of daily steps for the best of cardiometabolic health. 
First, as indicated by the results in Section \ref{subsubsec:global} on 90-day average daily step counts, the learned policy suggests people take similar numbers of daily steps consistently every 90 days, centering around 10,000 steps per day, rather than following an inconsistent and highly variable pattern of daily steps. The suggestion by the learned policy is aligned with commonly used step-count targets and step-based public-health indices \citep{TudorLocke2004StepsEnough,TudorLocke2011AdultsStepsEnough}.

Moreover, as indicated by the results in Section \ref{subsubsec:global} on the distribution of daily steps of 90 days within covariate subgroups, the learned policy never suggests increasing daily steps uniformly; instead, it recommends distributional changes which vary over low-, moderate-, and high-activity periods. 
The learned policy always recommends people to take significantly more daily steps than their behaviors during their high-activity periods of 90 days. For some subgroups, it also suggests more daily steps during their moderate-activity periods. The learned policy rarely suggests significant changes in daily steps during the low-activity periods, except for three subgroups of BMI: Normal, BMI: Obese, and Blood Pressure: Normal. 

The results of the All of Us data analysis are scientifically interpretable. For example, for participants with high baseline glucose, the recommended moderate-activity range expands upward (from approximately 5,471--7,597 to 6,629--11,861 steps), and the high-activity threshold increases to above 11,861 steps (Figure~\ref{fig:aou_glucose}). For obese participants defined by BMI, upward shifts occur across all activity periods, with the high-activity threshold increasing from above 8,701 to above 13,853 steps (Figure~\ref{fig:aou_bmi}). These patterns are consistent with the notion that individuals with less favorable cardiometabolic profiles may benefit from more substantial increases in PA \citep{Gay2016,Jayedi2024}.

\section{Discussion}\label{sec:discussion}

In this paper, we study precision PA prescription based on the daily step data from the All of Us Research Program. Instead of reducing PA to a scalar summary such as average daily steps, we represent the PA prescription as the conditional distribution of daily steps over 90 days given covariates, which can account for not only the amount of PA, but also how PA is distributed over 90 days. Methodologically, we propose to learn the optimal precision PA prescription in the form of a conditional distribution by offline RL where the action is a function. 
We develop two algorithms which extend fitted Q-evaluation and fitted Q-iteration respectively to the functional-action setting. The simulation results show that the proposed method outperforms representative offline RL methods that use an aggregated continuous action, suggesting the benefit of preserving the functional structure of the action. The application of our proposed method to 
the All of Us data provides reasonable and practical insights on precision prescriptions of daily steps for the best of cardiometabolic health based on blood glucose, BMI, blood pressure, age, and sex. The learned policy generally suggests people follow a more consistent and stable pattern of taking daily steps over time and also take more steps than their behaviors during their high-activity periods of 90 days.

This paper has several limitations which require further studies. 
First, the All of Us data used to learn the proposed precision PA prescription are observational data with a limited sample size.  
Thus intervention trials with a much larger sample size will be needed to study causation and to fit more complex policies. Moreover, the reward specification in this paper is based on cardiometabolic biomarkers, age, and sex. It is of interest to consider additional covariates and other health biomarkers, if available, to learn PA prescriptions to reduce other health risks.

\section*{Data Availability Statement} 
Data that support the findings of this study are available from the All of Us Research Program (Registered Tier Dataset v8) under a data use agreement. Restrictions apply to the availability of these data, which were used under license for this study. Data are available at https://www.researchallofus.org/ with the permission of the All of Us Research Program.

\section*{Conflict of Interest}
There are no financial or non-financial competing interests to report.

\section*{Acknowledgments}

In preparation of the manuscript, the authors used ChatGPT (GPT-5.5) to polish the
paper and check grammatical errors. The ChatGPT is only used for English editing. It is not used for any
other part of the paper.

\section*{Supplementary Material}

The Supplementary Material contains the following sections. Section S1 provides selected summary statistics of All of Us data. Section S2 presents the fitted-Q iteration algorithm for continuous actions. Section S3 details hyper-parameter selections. Section S4 provides the simulation study. Section S5 contains descriptive statistics for the discretized covariate subgroups used in the subgroup comparisons in Section \ref{sec:pa_prescription}.

\bibliographystyle{abbrv}

\bibliography{Bibliography-MM-MC-cleaned-JASA-v3}

\end{document}